\documentclass[aps,prd,twocolumn,groupedaddress]{revtex4-1}

\usepackage{graphicx}
\usepackage{hyperref}
\begin{document}


\title{Constraints to a Galactic Component of the Ice Cube cosmic neutrino flux from ANTARES}


\author{M. Spurio}  
\affiliation{Dipartimento di Fisica dell'Universit\`a di Bologna  \\
Istitito Nazionale di Fisica Nucleare - Sezione di Bologna \\ 
Viale Berti Pichat 6/2 - 40127 Bologna (Italy)}
\email[]{maurizio.spurio@bo.infn.it}



\begin{abstract}

The IceCube evidence for cosmic neutrinos in the high-energy starting events (HESE) sample has inspired a large number of hypothesis on their origin, mainly due to the poor precision on the measurement of the direction of showering events.
The fact that most of HESE are downward going suggests a possible Galactic component. This could be originated either by a single point-like source or to a directional excess from an extended Galactic region. 
These hypotheses are reviewed and constrained, using the present available upper limits from the ANTARES neutrino telescope.

ANTARES detects $\nu_\mu$ from sources in the Southern sky with an effective area larger than that providing the IceCube HESE for $E_\nu<60$ TeV and a factor of about two smaller at 1 PeV.
The use of the $\nu_\mu$ signal enables an accurate measurement of the incoming neutrino direction.
The Galactic signal allowed by the IceCube HESE and the corresponding ANTARES limits are studied in terms of a power law flux $E^{-\Gamma}$, with spectral index $\Gamma$ ranging from 2.0 to 2.5 to cover most astrophysical models.

\end{abstract}

\pacs{95.85.Ry, 95.55.Vj}

\maketitle

\section{Introduction}\label{intro}

The IceCube Collaboration announced in \cite{ic1} evidences for the first detection of extraterrestrial high-energy neutrinos using two years of data with the full detector, recently updated with a third year \cite{ic2}.
The estimated energies of events in the IceCube sample (high-energy starting events, HESE) range from 30 TeV to 2 PeV.
In the IceCube papers the hypothesis of a neutrino flux with flavor ratios $\nu_e:\nu_\mu:\nu_\tau= 1:1:1$ is considered.
This flux is exactly that expected from charged pion decays in cosmic ray (CR) accelerators and neutrino oscillation on their way to the Earth.
The non-observation of events beyond 2 PeV suggests a break or an exponential cutoff in the neutrino flux for a power law $\Phi(E) \propto E^{-\Gamma}$ and a hard spectral index, as for instance $\Gamma \simeq 2.0$. 
An unbroken power law is also compatible with the data assuming a softer spectrum, such as $\Gamma=2.3$.
The majority of the events are downward going; as the detector is located at the South Pole, this hints an important Galactic contribution. Some events have however high Galactic latitudes, indicating at least some extragalactic component.

Different implications of this discovery have been discussed widely.

A suppression in the PeV energy range, if present, is not typical for many models of extragalactic neutrino sources.
The maximal energy of neutrinos produced via pion production in proton-photon (p$\gamma$) or proton-gas (pp) interactions is approximately 5\% of the energy of the proton primary. 
If the HESE excess arises totally from isotropically distributed extragalactic sources, the corresponding all-sky neutrino diffuse flux is very close to the upper bound up to $\sim 10^{15}$ eV   \cite{wb}.
If Active Galactic Nuclei or $\gamma$-ray bursts were sources of ultrahigh energy cosmic rays, then they should produce neutrinos with energies up to $E\sim 10^{19}$ eV \cite{mesze}. 
The same holds for the cosmogenic neutrino flux which peaks around $10^{18}$  eV for proton primaries \cite{sta}. 

The observed HESE energy spectrum requires strong energy loss processes of protons up to 50 PeV that are not expected to take place within $\gamma$-ray bursts or AGN sources. 
A possible explanation \cite{wax} is that protons produce pions not within the sources, but in the environment surrounding them, as in starburst galaxies. These galaxies act as calorimeters: protons with energies $< 100$ PeV lose all their energy into pions after escaping the source that produced them but before escaping the galaxy. 
This mechanism could also explain the suppression of HESE above 2 PeV, as protons of energy exceeding 100 PeV may escape the galaxy without producing pions.
{In \cite{mura} it is shown that pp interactions in galaxy groups/clusters and star-forming galaxies can explain the IceCube events. The spectral index $\Gamma$ is constrained using the observed diffuse $\gamma$-ray background to be $\Gamma\le 2.1-2.2$ and $\Gamma=2.3$ is ruled out in pp scenarios. }
In an alternative model (see \cite{ste} and references therein), very high energy neutrinos could be produced in the cores of AGN, with a flux peaked at PeV energies. Protons accelerated by shocks in the vicinity of the black hole accretion disk and trapped by the magnetic field, lose energy dramatically by interactions with the dense photon field of thermal emission from the accretion disk.

An attempt to associate plausible astronomical counterparts in the GeV - TeV $\gamma$-rays to individual IceCube events was done in \cite{pr}. 
Here, sources in the available catalogues within the error circles of the IceCube events were looked for. The spectral energy distribution (SED) of these sources were also built and compared with the energy of the corresponding neutrino. 
The authors found that likely counterparts include mostly BL Lacs and two Galactic pulsar wind nebulae. 
However, the SED necessary to explain the neutrino production for most objects cannot be smoothly connected with the extrapolated SED obtained from $\gamma$-ray observations.

{A guaranteed component for the Galactic neutrino flux is due to the CRs interacting with gas during their confinement and different models computing the neutrino yields exist. These models rely on a variety of assumptions that include the CR and matter density in the Galaxy and the strength and orientation of the Galactic magnetic fields. Some of them \cite{inge,candia1} assume an isotropic density of CRs as well as of the interstellar matter. Models that are more realistic take diffusion and drift of CRs in the Galaxy into account \cite{candia2}. Larger neutrino fluxes from the Galactic center region are obtained when the local enhancement in the CR density in this region, where also the matter density is maximal, is accounted for. }

{In most computations, the expected neutrino flux from such mechanism is marginal with respect to the measured flux of HESE events. In \cite{tcher} it was derived that more than 20 years are necessary for IceCube to do a detection. Another calculation of the neutrino yield due to the propagation of CRs in the Galaxy which takes into account the CR elemental composition is reported in \cite{joshi}. Here, it is concluded that at most 0.1 of the observed HESE in IceCube can be attributed to CR interactions with matter.
In a third work \cite{1405.3797}, even in the direction of the largest expected intensity, the prediction is about two orders of magnitude too small to explain the IceCube excess. Moreover, due to the very slim Galactic plane, the events should be concentrated within $|b| \le 1^\circ$ which is much narrower than the latitude distribution of the IceCube events.
Finally, in \cite{guo} diffuse Galactic neutrinos may contribute to a fraction of the HESE below $\sim 100$ TeV. The high-energy events, however, require another origin with harder spectral index than that foreseen by this mechanism. }

{A possible explanation of the brightest spots in the Galactic neutrino sky is that an enhanced neutrino production might occur in giant molecular clouds immersed into the CR over-densities close to recent CR sources \cite{1405.3797}. 
A neutrino yield, assuming an unbroken power-law spectrum from optically thin Galactic neutrino sources, is considered in \cite{Anchordoqui:2013qsi}.
Here, it is found that the hypothesized unbroken power-law spectrum with spectral index  $\Gamma =2.3$ is consistent at the 1.5 standard deviation level with the observed HESE up to 2 PeV.  }

A contribution from point-like Galactic sources is suggested by the presence of clustering of events with errors compatible with the emission from a single source in the sky, and the presence of a \textit{hot spot} near the Galactic Center. 
A fraction of the IceCube HESE could arise from the \textit{Fermi bubbles} region \cite{fb} or from the Galactic halo \cite{bai,tay}. 
Different authors (\cite{ahlers,razza} and references therein) have considered the  TeV and PeV $\gamma$-ray upper limits placed by Fermi-LAT and EAS detectors. 
These observations also motivate the hypothesis that the IceCube excess could originate from a restricted Galactic region. 
The shape of the $\gamma$-ray and neutrino spectra originated from the interaction of CR protons with ambient protons for sources located in the Galactic Center region was considered in detail in \cite{supa}.
In \cite{nero} it was suggested that both the IceCube HESE and the diffuse $\gamma$-ray emission from the Galactic Plane measured by Fermi-LAT are produced in interactions of CRs with the interstellar medium in the Norma arm and/or in the Galactic Bar. 
CRs responsible for the $\gamma$-ray and neutrino flux are characterized by a hard spectrum with the slope harder than $E^{-2.4}$ and cutoff energy higher than 10 PeV.

Finally, the neutrino excess has been associated, e.g., with unidentified TeV $\gamma$-ray sources \cite{fox} or dark matter emission mechanisms \cite{ser,zav}. 

In this paper models involving a Galactic origin of part of the IceCube HESE, either from point-like sources or from diffusion processes in restricted regions of the Galaxy, are discussed.
In both cases, the corresponding neutrino flux could produce a signal in the ANTARES neutrino telescope located in the Mediterranean Sea. 
The upper bounds already set by ANTARES for sources located in the Southern hemisphere are used to constrain Galactic models based on the IceCube neutrino sample.

The response of a detector in terms of collected neutrinos depends on the neutrino effective area $A_{eff}(E)$. 
The neutrino effective area is a strong function of the neutrino energy, and the number of detected events depends on the assumed energy spectrum from sources. 
In section \ref{sez:aeff}, the effective area yielding the IceCube HESE is compared with that used by ANTARES in the search for cosmic neutrino sources.

The standard diffusive shock acceleration yields a $\Gamma=2.0$ spectral index for primary CRs, and thus for secondary $\gamma$-rays and neutrinos \cite{spu}. 
However, most $\gamma$-ray sources observed in the GeV and TeV range show spectral indexes larger than 2.0.  
{Simple models of CR interactions with gas do not lead to any expected softening of the $\gamma$-ray spectra. Indeed, the reason why $\gamma$-ray spectra from supernovae remnants are observed with spectral indexes $\Gamma\simeq 2.2-2.3$  remains unclear. }
This motivated in section \ref{sez:IC} the study of the IceCube signal, done using the published effective area, in terms of a power-law flux $E^{-\Gamma}$, with spectral indexes $\Gamma$ ranging from 2.0 to 2.5.  
The fraction of the IceCube HESE signal that can be of Galactic origin is quantitatively estimated in section \ref{sez:HESEga}.

Sources located in the Galactic region can be studied by the ANTARES detector through the $\nu_\mu$ charged current (CC) interactions with a much higher angular precision and with an effective area comparable to that of the IceCube HESE sample. 
The 90\% C.L. upper limits set by ANTARES in the Galactic Plane \cite{anta} for a $E^{-2}$ flux are translated in section \ref{sez:anta} to limits for different spectral indexes $\Gamma$ using the ANTARES effective area. 
In section  \ref{sez:GalIC} these upper bounds are used to constrain models in which a fraction of the IceCube signal originates from Galactic point-like sources.

In section \ref{Galdiff}, the intensity of an enhanced flux of high-energy neutrinos from a restricted Galactic region compatible with the IceCube HESE is evaluated assuming different spectral indexes $\Gamma$.
The level of expected sensitivities from the ANTARES telescope to observe a directional excess according to different hypotheses are considered. 
To test the hypothesis, a dedicated analysis similar to the that done to set limits on the neutrino flux from the Fermi bubbles region \cite{anta-fb} is required.
The results are discussed in section \ref{sez:disc}, as well as the perspectives for the future KM3NeT neutrino telescope in the Mediterranean Sea.

\section{Effective areas for $\nu$ telescopes in the Southern and Northern hemisphere}\label{sez:aeff}

\begin{figure*}[tb]
\begin{center}
\includegraphics[width=11.0cm]{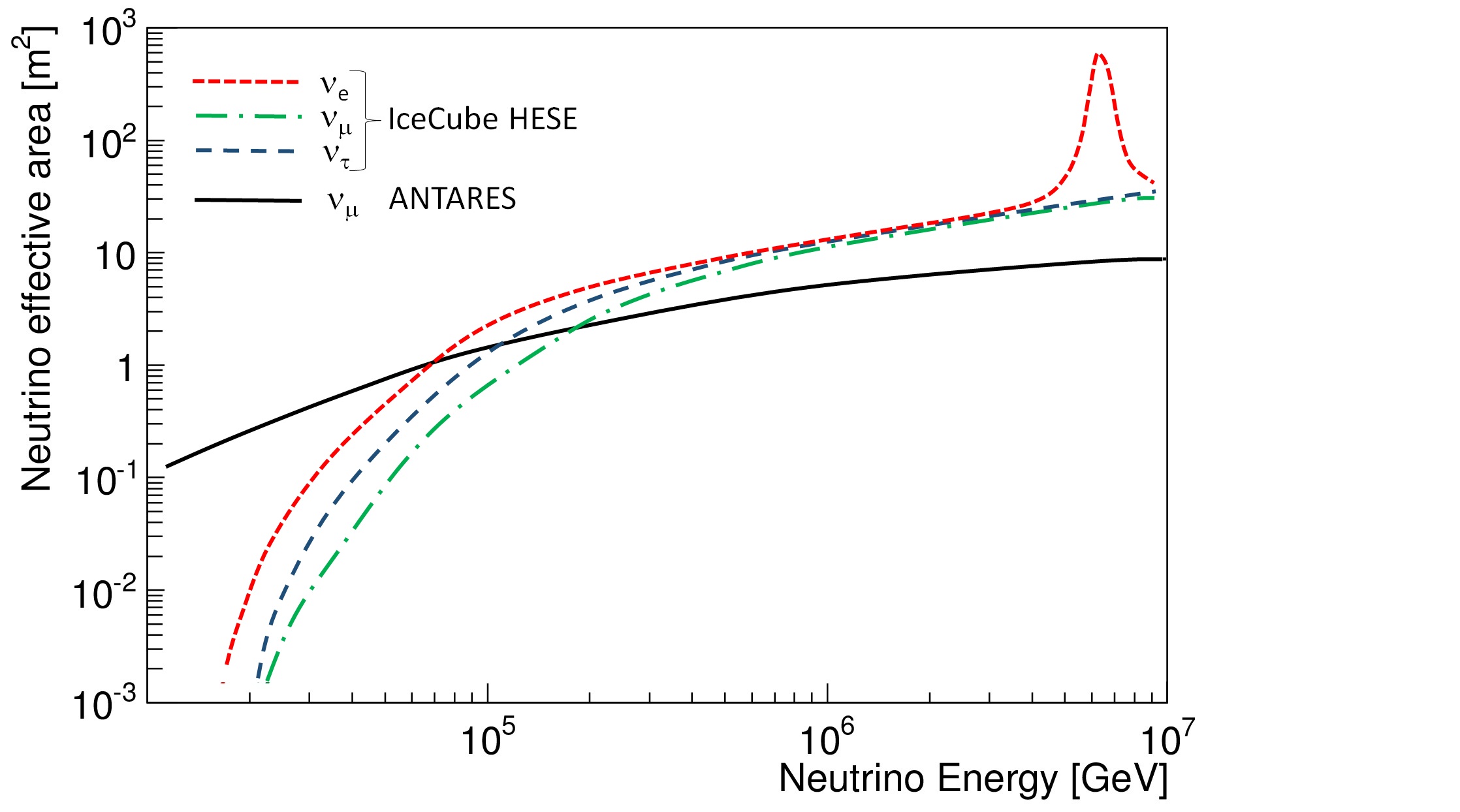}
\end{center}
\caption{\small\label{fig:Aeff} The red (dashed), green (dot-dashed) and blue (long-dashed) lines refer to $\nu_e, \nu_\mu$ and $\nu_\tau$ neutrino effective area reported in the Ice Cube analysis providing the first evidence for a high-energy neutrino flux of extraterrestrial origin \cite{ic1}. 
The full black line refers to the ANTARES effective area for the $\nu_\mu$ flavor obtained in the search for point-like sources \cite{anta12}. 
The effective area depends on the cuts of the selection analyses. Event rates can be obtained by folding the assumed neutrino spectrum with the effective areas (see text).}
\end{figure*}

The neutrino effective area at a given energy, $A_{eff}(E)$, is defined as the ratio between the neutrino event rate (units: s$^{-1}$) in a detector and the neutrino flux (units: cm$^{-2}$ s$^{-1}$) at that energy. 
The effective area depends on the flavor and cross-section of neutrinos, on their absorption probability during the passage through the Earth, and on detector-dependent efficiencies.
Detector efficiencies are correlated to each particular analysis, referring to the criteria used to trigger and to reconstruct the event, and to the quality cuts applied to reduce the background. 

Figure \ref{fig:Aeff} shows the effective area of the two running neutrino observatories in the Antarctic and in the Mediterranean Sea as a function of the neutrino energy.
The red ($A_{eff}^{\nu_e}$), green ($A_{eff}^{\nu_\mu}$) and blue ($A_{eff}^{\nu_\tau}$) lines refer to the IceCube HESE \cite{ic1,ic2} and are valid over the whole solid angle. 
The black line corresponds to the ANTARES effective area for the $\nu_\mu$ flavor as derived in the analysis \cite{anta12} for the search for cosmic neutrino point sources in the declination band containing the Galactic Center. 
For all considered effective areas, the flux for a given flavor is assumed to consist of equal amounts of $\nu$ and $\overline \nu$.


The ANTARES effective area shown in Fig. \ref{fig:Aeff} is larger than that of IceCube HESE (irrespective of the neutrino flavor) at energies below $\sim$ 60 TeV, despite the fact that its instrumented volume is much smaller than 1 km$^3$.
{At 100 TeV (1 PeV), the ANTARES effective area for the $\nu_\mu$ flavor is a factor of 1.7 (0.45) that of IceCube $A_{eff}^{\nu_\mu}$. When ($A_{eff}^{\nu_e}$+$A_{eff}^{\nu_\mu}$+$A_{eff}^{\nu_\tau}$) is considered, the IceCube HESE effective area is a factor of 3.5 (7.3) larger than that of ANTARES for the $\nu_\mu$ flavor at 100 TeV (1 PeV).}

As mentioned, the IceCube and ANTARES effective areas $A_{eff}$ include the selection efficiencies for the particular analyses.
Large volume neutrino detectors suffer from a huge background of downward-going atmospheric muons \cite{mupara}.
The IceCube selection criteria necessary for a $4\pi$ sr selection of cosmic neutrinos reduce the effective area due to the request that the vertex of the $\nu$ interaction occurs inside a restricted detector volume. 
In addition, a large deposited energy in the instrumented volume is required. 
These criteria, as explained in the \textit{materials and methods} section of \cite{ic1}, are able to largely suppress the background induced by atmospheric muons and by atmospheric neutrinos.
On the other hand, these requirements induce a suppression of $\nu_\mu$ candidates, inhibiting the detection of neutrino-induced muons with vertices outside the instrumented volume.

The ANTARES effective area refers to a completely different detection strategy. Only upgoing muons are selected, and no restriction for a confinement volume for the vertex interaction is imposed. The reported ANTARES effective area is for the $\nu_\mu$ flavor, and limited to upgoing neutrinos.  
Only atmospheric neutrinos and a small fraction of wrongly reconstructed atmospheric muons (at the level of $\le$ 10\% of atmospheric neutrinos) constitute the background to the cosmic signal.   

\section{The normalization factors for a diffuse HESE signal}\label{sez:IC}
Table \ref{tab:ICud} reports the number of data, background and signal events in the energy range of 60 TeV$< E_{dep}<$ 3 PeV, for the HESE in the whole solid angle and for the upgoing and downgoing sub-samples \cite{ic2}.
\begin{table}[tbh]
{\centering \begin{tabular}{l|c|c|c||c}
\hline 
& Data & Bck & $n^\prime _{IC}$ & $N^\prime_{IC}$ \\
&      &     &          & $E^{-2}$ (best fit) \\ \hline \hline 
All       	& 20	&2.7	&17.3	&18.2		\\ \hline 
Up (North)	& 5	&1.4	& 3.6	&6.7	\\
Down (South) &15	&1.3	&13.7	&11.5	\\
\hline
\end{tabular}\par}
\caption{{\small Number of HESE between 60 TeV$< E_{dep}<$ 3 PeV (column 2), estimated number of background events (column 3) and number $n^\prime _{IC}$  of signal events (column 4) for the whole sample (All), and separated for the fraction of upgoing (from the Northern hemisphere) and downgoing  (from the Southern hemisphere) signal. The last column report the expected signal $N^\prime _{IC}$ in the same energy range assuming a $E^{-2.0}$ power law with the normalization factor reported in Eq. (\ref{eq:ic0}).
\label{tab:ICud}}}
\end{table}


The best-fit astrophysical flux per-flavor (e.g., $\nu_\mu+\overline\nu_\mu$) for a $E^{-2}$ spectrum in the energy interval 60 TeV - 3 PeV based on the 17.3 events in the $4\pi$ sr is \cite{ic2}:
\begin{eqnarray}\label{eq:ic0}
E^2 \Phi(E) &\equiv& \Phi^{D,2.0}_0 \nonumber \\
&=& (0.95\pm 0.3)\times 10^{-8}  \textrm{ GeV cm}^{-2} \textrm{s}^{-1}\textrm{sr}^{-1}
\end{eqnarray}
The above notation refers to the fact that a diffuse ($D$) flux is considered, with a spectral index $\Gamma=2.0$. The quantity $\Phi^{D,2.0}_0$ represents the derived normalization factor.

The $E^{-2}$ model describes the data well; however, if unbroken, this spectrum predicts 3.1 additional events above 2 PeV, which are not observed.
This may indicate either a softer spectrum or a cutoff at high energies.
Assuming a softer spectrum and no cutoff, the best-fit power law for $E_\nu>60$ TeV as reported by IceCube (Fig. 4 of \cite{ic2}) is:
\begin{equation}\label{eq:ic}
E^{2} \Phi(E) = 1.5\times 10^{-8} \ (E/\Lambda)^{-0.3} \textrm{ GeV cm}^{-2} \textrm{s}^{-1}\textrm{sr}^{-1} 
\end{equation}
where $\Lambda\equiv 100\textrm{ TeV}$ is a scale factor with the dimension of an energy.
The above relation is thus equivalent to:
\begin{equation}\label{eq:ic-br1}
{E^{2.3}\over \Lambda^{0.3} } \Phi(E) = 1.5\times 10^{-8} \  \textrm{ GeV cm}^{-2} \textrm{s}^{-1}\textrm{sr}^{-1} \ .
\end{equation}
In some cases, in the literature (for instance, in \cite{Anchordoqui:2013qsi,ancho}) the scale factor is assumed to be $\lambda \equiv$ 1 GeV, and thus Eq. (\ref{eq:ic-br1}) is equivalent to:
\begin{eqnarray}\label{eq:ic1}
{E^{2.3}\over \lambda^{0.3} }  \Phi^{D,2.3}(E) &\equiv& \Phi^{D,2.3}_0 \nonumber \\ 
&=& 4.7\times 10^{-7}   \textrm{ GeV cm}^{-2} \textrm{s}^{-1}\textrm{sr}^{-1}  .
\end{eqnarray}
Usually, the $\Lambda,\lambda$ are omitted in writing the equations.
The normalization factor $\Phi^{D,2.3}_0 $ defined above represents the neutrino flux per flavor at 1 GeV. In the following, the convention of Eq. (\ref{eq:ic1}) is adopted, omitting to make explicit $\lambda$.

More generally, the normalization factors $\Phi^{D,\Gamma}_0$ (in units: GeV cm$^{-2}$ s$^{-1}$ sr$^{-1}$) for a cosmic flux ${E^\Gamma \over \lambda^{\Gamma-2}} \Phi^{D,\Gamma} (E) \equiv
{E^\Gamma } \Phi^{D,\Gamma} (E)$ can be derived for a generic spectral index $\Gamma$.
Our approach is to obtain the same number $N_{IC}$ of expected events evaluated using a $E^{-2}$ spectrum with the normalization factor (\ref{eq:ic0}) when using the effective areas of Fig. \ref{fig:Aeff}.
In fact for an isotropic neutrino flux $\Phi^{D,\Gamma} (E) $ in a detector livetime $T$ and effective area $A_{eff}(E)\equiv [A_{eff}^{\nu_e}+ A_{eff}^{\nu_\mu}+ A_{eff}^{\nu_\tau}]$, the $N_{IC}$ is given by:
\begin{eqnarray}\label{eq:ic2}
N_{IC} &=& T \cdot \int \Phi^{D,\Gamma} (E) \cdot [A_{eff}^{\nu_e}+ A_{eff}^{\nu_\mu}+ A_{eff}^{\nu_\tau}] \cdot dE\cdot d\Omega \nonumber \\
    &=& 4\pi T \cdot \Phi^{D,\Gamma}_0 \cdot \int   E^{-\Gamma} \cdot A_{eff}(E)\cdot dE \nonumber \\
&=& 4\pi T \cdot \Phi^{D,\Gamma}_0 \cdot {\cal D}_\Gamma \ .
\end{eqnarray}
The integral, denoted as ${\cal D}_\Gamma$, extends over the energy range where $ A_{eff}(E)$ is not null, and can be computed numerically. 
${\cal D}_\Gamma$ corresponds to the detector response  for a given energy spectrum $E^{-\Gamma}$. 
The value $N^\prime_{IC}$ reported in the last column of Table \ref{tab:ICud} refers to the integration in the energy interval between 60 TeV and 3 PeV.

Table \ref{tab:IC} reports the normalization factors $\Phi^{D,\Gamma}_0$ computed using Eq. (\ref{eq:ic2}) for $\Gamma\ge 2.0$ and assuming an unbroken spectrum.
Note that as the value of $\Gamma$ increases (softer spectrum), the normalization factor increases. 
This occurs because the effective area decreases as the neutrino energy decreases.
In our computation, the normalization factor for the spectral index $\Gamma=2.3$ corresponds to $\Phi^{D,2.3 }_0=4.6\times 10^{-7} \textrm{ GeV cm}^{-2} \textrm{s}^{-1}\textrm{sr}^{-1}$, to be compared with the value (\ref{eq:ic1}) obtained by the fit on the $E_{dep}$ performed by IceCube in \cite{ic2}.
\begin{table}[tbh]
{\centering \begin{tabular}{l|c|c||c|c}
\hline 
$\Gamma=$ & $\Phi^{D,\Gamma}_0 $ & ${\Phi^{D,\Gamma}_0\over  \Phi^{D,2}_0}$ & 0.1-1 PeV & $>1$ PeV \\ 
& (GeV cm$^{-2}$ s$^{-1}$ sr$^{-1}$) & & & \\ \hline
2.0 & $0.95\times 10^{-8}$ & 1 & 60\% & 31\%\\ 
2.1 & $3.5\times 10^{-8}$ & 3.7 & 63\% & 26\%\\ 
2.2 & $1.2\times 10^{-7}$ & 13.5 & 66\% & 21\%\\ 
2.3 & $4.6\times 10^{-7}$ & 48 & 68\% & 17\%\\ 
2.4 & $1.6\times 10^{-6}$ & 170 & 68\% & 13\% \\ 
2.5 & $5.4\times 10^{-6}$ & 580 & 68\% & 10\%\\ 
\hline
\end{tabular}\par}
\caption{{\small Normalization factors for the IceCube HESE, assuming a diffuse flux and different spectral indexes $\Gamma$. The third column reports the ratio with respect to the normalization factor given in Eq. (\ref{eq:ic0}) and obtained for the $E^{-2}$ spectrum. The last two columns report the percentage of the IceCube HESE induced by cosmic neutrinos in the energy range 100 TeV- 1 PeV and above 1 PeV evaluated using the detector response  ${\cal D}_\Gamma$, defined in Eq. (\ref{eq:ic2}).
\label{tab:IC}}}
\end{table}

The values reported in Table \ref{tab:IC} are in agreement with that obtained with a different method in \cite{Anchordoqui:2013qsi,ancho}, where information included in \cite{ic1} are used.
The conclusions reported in \cite{ancho} is that, given the current statistics, it is possible to characterize the IceCube neutrino energy spectrum with a single power law, and the $\Gamma=2.3$ was considered as the reference value for the unbroken power law. 

Using the response function ${\cal D}_\Gamma$, it is possible to evaluate the number of induced signal events in a given energy interval $E_{min}$ - $E_{max}$. In the last two columns of Table \ref{tab:IC} the fractions of the IceCube HESE signal from 100 TeV to 1 PeV and above 1 PeV are reported, summing the contribution for an isotropic $\nu_e, \nu_\mu,\nu_\tau$ flux.

\section{A galactic component in the HESE sample\label{sez:HESEga}}

No hypothesis test on HESE events \cite{ic2} yielded statistically significant evidence of clustering or correlations, in particular from the Galactic Center or Galactic Plane.
A Galactic component seems however present.
The last column of Table \ref{tab:ICud} reports the expected number of events in the energy range 60 TeV$< E_{dep}<$ 3 PeV for the $E^{-2}$ spectrum. Similar values are obtained for $\Gamma>2.0$ (e.g., 7.5 upgoing and 12.0 downgoing events for $\Gamma=2.3$). 
By comparing the observed and expected number of signal events in the Up/Down case in Table \ref{tab:ICud}, the expected signal from the Northern hemisphere is a factor of $\sim 2$ larger than measured, while the expected signal from the Southern hemisphere is smaller than measured.
This is true for all the considered values of $\Gamma$.

The Northern sky, yielding upgoing events in IceCube, contains a negligible fraction of the Galaxy. The 3.6 observed signal events can thus be considered purely of extragalactic origin. 
Taking into account the difference in the effective areas for upgoing and downgoing events, the expected upgoing HESE fraction is 37\% (39\%) for a $E^{-2.0}$ ($E^{-2.5}$) spectrum.
Assuming 3.6 upgoing events and a symmetric contribution from the extragalactic sky, the expected downgoing HESE of extragalactic origin are 6.2 (5.8) events for the $E^{-2.0}$ ($E^{-2.5}$) spectrum.
The measured downgoing signal corresponds to 13.7 events, with an excess of 7.5 (7.9) events with respect to the above hypothesis. 
This excess could be attributed to the presence of the Galaxy in the Southern sky.

In the following sections, the normalization factor given by Eq. (\ref{eq:ic0}) (with quoted uncertainty of $\pm 30\%$) is used , with a possible Galactic contribution up to 6 events out of the 17.3 HESE signal in the 60 TeV $< E_{dep} <$ 3 PeV range.
The normalization factor that reproduces the number of downgoing events 
is $\sim 20\%$ higher than Eq. (\ref{eq:ic0}). 
The largest group of events spatially clustered (the \textit{hot spot}) corresponds (see Suppl. Table V in \cite{ic2}) to 6 events near the Galactic Center.  

\section{ANTARES constraints for the Galactic Center region}\label{sez:anta}

A dedicated search for an excess of events of cosmic origin was done by the ANTARES collaboration \cite{anta} in a wide region of $20^\circ$ around the hot spot ($\alpha=-79^\circ, \delta=-23^\circ$) reported in \cite{ic1}.
A signal over the background of atmospheric neutrinos was searched for, with the assumptions of a point-like source and of three Gaussian-like source extensions of $0.5^\circ, 1^\circ$ and $3^\circ$ diameter. 
By definition, a point-like source has an extension that cannot be resolved by a detector. 
In the considered ANTARES sample, a point-like source cannot be resolved if its angular size is smaller than $\theta^A_{res} \sim 0.4^\circ$. 
Data collected from 2008 to 2012 were analyzed using the full-sky algorithm with the likelihood presented in \cite{anta12}.
No significant cluster was found. 

The flux from a point-like source, or from a restricted region in the sky, is expressed in units of $\textrm{ GeV cm}^{-2} \textrm{s}^{-1}$.
The derived ANTARES upper limits for a $E^{-2}$ signal are slightly dependent on the declination assumed for the source and change by about $ \pm 20\%$ in the band $-40^\circ <\delta<-5^\circ$ (Figure 4 of \cite{anta}).  In the following, the values corresponding to the declination of the Galactic Center: 
\begin{eqnarray}\label{eq:ant1}
E^2\Phi^{A,2.0}\equiv \Phi^{A,2.0}_0 &\simeq& 4.0 \times 10^{-8} \textrm{ GeV cm}^{-2} \textrm{s}^{-1} \textrm{, point-like} \nonumber \\ 
&\simeq& 5.0\times 10^{-8} \textrm{ GeV cm}^{-2} \textrm{s}^{-1} 
\textrm{ , for } 0.5^\circ \nonumber \\ 
&\simeq& 6.5\times 10^{-8} \textrm{ GeV cm}^{-2} \textrm{s}^{-1} \textrm{ , for } 1^\circ \nonumber \\ 
&\simeq& 10\times 10^{-8} \textrm{ GeV cm}^{-2} \textrm{s}^{-1} \textrm{ , for } 3^\circ 
\end{eqnarray}
are considered.

These 90\% C.L. upper limits rely on the assumption that a given number of signal events, $n^A_s$, are produced by a $E^{-2}$ signal spectrum. 
The equivalent 90\% C.L. upper limits for any other spectral index $\Gamma$ can be derived with a procedure similar to that used for the IceCube signal. 
In fact, $n^A_s$ is obtained by the integration of a given cosmic neutrino spectrum with the ANTARES effective area $ A^A_{eff}(E) $ in Fig. \ref{fig:Aeff}. 
To give a feeling about the atmospheric neutrino background, considering the total number of observed ANTARES events (5516) in the visible half sky hemisphere, on average 0.2 (0.8) events are expected in a cone of 0.5$^\circ$ (1$^\circ$).
Assuming a source spectrum 
$\Phi^{A,\Gamma}(E)= \Phi^{A,\Gamma}_0 E^{-\Gamma}$, $ n^A_s$ is given by:
\begin{eqnarray}\label{eq:ant2}
n^A_s &=& T^A \cdot \int {\Phi^{A,\Gamma}(E)}\cdot A^A_{eff}(E) \cdot dE \nonumber \\
&=&  T^A \cdot \int \Phi^{A,\Gamma}_0 E^{-\Gamma}\cdot A^A_{eff}(E) \cdot dE 
\end{eqnarray}
From this relation, using for $\Gamma=2.0$ the values reported in (\ref{eq:ant1}) (point-like case), the 90\% C.L. upper limits on the normalization factors $\Phi^{A,\Gamma}_0$ are obtained and reported in Table \ref{tab:ANTA}.
\begin{table}[tb]
{\centering \begin{tabular}{l|c|c||c|c}
\hline 
$\Gamma=$ & $\Phi^{A,\Gamma}_0$ (GeV cm$^{-2}$ s$^{-1}$)  & ${\Phi^{A,\Gamma}_0/  \Phi^{A,2}_0}$ & 0.1-1PeV & $>$ 1 PeV \\ \hline
2.0 & $4.0\times 10^{-8}$ & 1 & 36\% & 13\%\\
2.1 & $1.2\times 10^{-7}$ & 2.9 & 31\% & 9\%\\
2.2 & $3.2\times 10^{-7}$ & 8.1 & 26\% & 6\%\\
2.3 & $8.4\times 10^{-7}$ & 21 & 21\%  & 4\%\\
2.4 & $2.2\times 10^{-6}$ & 56 & 17\% & 2.6\%\\
2.5 & $5.5\times 10^{-6}$ & 138 & 13\% &1.6\%\\ \hline
\end{tabular}\par}
\caption{{\small 90\% C.L. upper limits (column 2) computed for a neutrino point-like source for different spectral indexes $\Gamma$ as obtained using Eq. (\ref{eq:ant2}), and their ratio (column 3) with respect to the case $\Gamma=2.0$. For $\Gamma=2.0$, the published ANTARES 90\% C.L. limit for a point-like source is used, Eq. (\ref{eq:ant1}); for source widths of $0.5^\circ, 1.0^\circ$ and $3.0^\circ$, the limits increase accordingly to Eq. (\ref{eq:ant1}).
The last two columns report the percentage of the ANTARES signal induced by cosmic neutrinos in the energy range 100 TeV- 1 PeV and above 1 PeV computed with the use of Eq. (\ref{eq:ant2}).
\label{tab:ANTA}}}
\end{table}

Similarly for the case of IceCube, for a softer spectrum the normalization factor increases because of the energy dependence of $A^A_{eff}$.
As the ANTARES effective area is larger than that of IceCube at low energies, the ratios ${\Phi^{A,\Gamma}_0/  \Phi^{A,2}_0}$ (column 3 of Table \ref{tab:ANTA}) are smaller than the corresponding ones of IceCube (column 3 of Table \ref{tab:IC}).
The fractions of the ANTARES signal between 100 TeV - 1 PeV and above 1 PeV are reported in the last two columns of Table \ref{tab:ANTA}.


\begin{table*}[tbh]
{\centering \begin{tabular}{c||c|c|c|c|c||c}
\hline 
&  \multicolumn{6}{c}{{ units: (GeV cm$^{-2}$ s$^{-1}$ sr$^{-1}$)  }} \\

& \multicolumn{5}{|c||}{{ $\Phi^{p,\Gamma}_0$ (from HESE) }} & ANTARES \\
$\Gamma=$ & $n_p=1$ &$n_p=2$ &$n_p=3$ &$n_p=4$ &$n_p=5$ & 90\% C.L. limit
 \\ \hline
2.0 & $6.9\  10^{-9}$ & $1.4\  10^{-8}$ & $2.1\  10^{-8}$ & $2.8\  10^{-8}$ &  {$3.5\  10^{-8}$} & $4.0\  10^{-8}$ \\
2.1 & $2.6\  10^{-8}$ & $5.1\  10^{-8}$ & $7.7\  10^{-8}$ &  {$1.0\  10^{-7}$} & \underline{$1.3\  10^{-7}$} & $1.2\  10^{-7}$ \\
2.2 & $9.0\  10^{-8}$ & $1.8\  10^{-7}$ & $2.7\  10^{-7}$ & \underline{$3.6\  10^{-7}$} & - & $3.2\  10^{-7}$ \\
2.3 & $3.3\  10^{-7}$ & $6.6\  10^{-7}$ & \underline{$9.9\  10^{-7}$} & - & - & $8.4\  10^{-7}$ \\
2.4 & $1.2\  10^{-6}$ & \underline{$2.3\  10^{-6}$} & - & - & - & $2.2\  10^{-6}$ \\
2.5 & $3.9\  10^{-6}$ & \underline{$7.9\  10^{-6}$} & - & - & - & $5.5\  10^{-6}$ \\
\hline
\end{tabular}\par}
\caption{{\small Column 2 to 6: normalization factors $\Phi^{p,\Gamma}_0$ for a point-like source producing $n_p=1,.. .,5$ IceCube HESE, for different values of the spectral index $\Gamma$. The last  column shows the 90\% C.L. upper limits for a point-like source derived from the ANTARES result assuming a $E^{-2}$ spectrum. The first value in each row excluded by the ANTARES upper limit is underlined.
\label{tab:point}}}
\end{table*}

\section{A point-like Galactic contribution to the IceCube signal}\label{sez:GalIC}

The IceCube HESE consist mostly of $\nu_e$ and $\nu_\tau$ from the Southern hemisphere producing showers in the detector. 
The energy released in these contained showers can be reconstructed at a level of 15\%, and their direction measured to within 10-15 degrees. 
Thus, the localization in the sky of event clustering can be determined with a poor precision. Most of the spatially clustered event groups (including the hot spot) are in the Southern hemisphere.
As discussed in section \ref{sez:aeff}, the detector located in the Northern hemisphere has an effective area large enough to test models of event clustering.

The incoming direction of muon neutrinos (1/3 of the total expected neutrino flux) can be much better measured using a different strategy, as the one used by ANTARES for the searches for point-like sources. 
The localization in the Mediterranean Sea ensures the visibility of declinations up to $\delta \sim +48^\circ$, including a large fraction of the Galactic Plane. 
Below 1 PeV, the presence of matter does not inhibit the arrival of neutrinos traversing the Earth, whose absorption effect is included in the computation of $A_{eff}$.
The upward-going muons produced in the $\nu_\mu$ CC interactions provide long tracks that can be reconstructed with sub-degree precision.

In this section, the number of IceCube HESE belonging to the same point-like source under the hypothesis of different energy spectra $E^{-\Gamma}$ is constrained using the ANTARES upper limits reported in Table \ref{tab:ANTA}.
In \cite{gonza}, for instance, a cluster of seven events out of the 28 of \cite{ic1} is predicted to correspond to a flux of $6\times 10^{-8}$ GeV cm$^{-2}$ s$^{-1}$ assuming an E$^{-2}$ spectrum, which was already ruled out in \cite{anta}.

Let us assume that a cluster of $n_p$ events out of $N_{IC}$ is produced by a point-like source located in the Galactic region. 
The normalization factor for the flux from a Galactic point-like source inducing $n_p$ events can be evaluated using the IceCube effective area, the reported excess (\ref{eq:ic0}) and assuming a source power-law spectrum:
\begin{equation}\label{eq:gal2}
E^\Gamma \Phi^{p,\Gamma} (E) = \Phi^{p,\Gamma}_0 \quad \textrm{ units: GeV cm}^{-2} \textrm{ s}^{-1} \ .
\end{equation}
The generic spectral index $\Gamma$ ranges from 2.0 to 2.5 in order to include most astrophysical models.
The normalization factor $\Phi^{p,\Gamma}_0$ necessary to produce $n_p$ events is obtained by requiring that:
\begin{eqnarray}\label{eq:gal3}
n_p &=& T \cdot \int \Phi^{p,\Gamma} (E) \cdot A_{eff}(E)\cdot dE \nonumber \\
    &=& T \cdot \int \Phi^{p,\Gamma}_0  E^{-\Gamma} \cdot A_{eff}(E)\cdot dE \nonumber \\
    &=& T \cdot \Phi^{p,\Gamma}_0  \int E^{-\Gamma} \cdot A_{eff}(E)\cdot dE \nonumber \\
    &=& T \cdot \Phi^{p,\Gamma}_0 \cdot {\cal D}_\Gamma
\end{eqnarray}
where the livetime $T$, the effective area $ A_{eff}(E)$ and, consequently, the detector response ${\cal D}_\Gamma$ are the same as in Eq. (\ref{eq:ic2}). 
This represents the fact that the $n_p$ events belong to the IceCube HESE sample.
Then, deriving the detector response from (\ref{eq:ic2}) and inserting into (\ref{eq:gal3}):
\begin{equation}\label{eq:gal4}
n_p = {N_{IC} \cdot \Phi^{p,\Gamma}_0  \over  4\pi \cdot \Phi_0^{D,\Gamma} }
\end{equation}
and, correspondingly, the normalization factor for a point-like source flux of a given spectral index $\Gamma$:
\begin{equation}\label{eq:gal5}
\Phi^{p,\Gamma}_0 = 4\pi\cdot \biggl({n_p \over N_{IC}}\biggr) \cdot \Phi_0^{D,\Gamma} \ .
\end{equation}
Table \ref{tab:point} reports the normalization factor $\Phi^{p,\Gamma}_0$ for a point-like source necessary to produce $n_p=1\div 5$ among the observed IceCube neutrino events.

The last column of Table \ref{tab:point} reports the 90\% C.L. upper limits derived from the ANTARES data. 
A point-like source near the Galactic Center yielding a cluster of more than 5 IceCube HESE  is excluded for a spectral index $\Gamma=2.0$. 
A source with an intensity able to produce more than two events is excluded for $\Gamma=2.3$.
The minimum number of events in a cluster already excluded by ANTARES for a given value of $\Gamma$ is underlined. 

As all values based on the normalization factor $\Phi^{D,2.0}_0 $ reported in (\ref{eq:ic0}), the uncertainties on the $\Phi^{p,\Gamma}_0$'s are at least $\pm 30\%$.
The ANTARES upper limits vary by $\pm 20\%$ with respect to the values reported in Table \ref{tab:point}, depending on the declination. 
If the source is assumed to have extension of $0.5^\circ, 1^\circ$ and 3$^\circ$ the ANTARES upper limits increase, scaling as reported in Eq. (\ref{eq:ant1}).

\section{A diffuse Galactic contribution to the IceCube signal}\label{Galdiff}

\begin{table*}[tbh]
{\centering \begin{tabular}{c|c||c|c|c|c||c}
\hline 
& & \multicolumn{5}{c}{{ units: (GeV cm$^{-2}$ s$^{-1}$ sr$^{-1}$)  }} \\
$\Delta \Omega$ & & \multicolumn{4}{|c||}{{ $\Phi^{D^\prime,\Gamma}_0$ (from HESE) }} & ANTARES \\
(sr) & $\Gamma=$ & $ n_{\Delta \Omega}=3$ & $ n_{\Delta \Omega}=4$ &$ n_{\Delta \Omega}=5$ &$ n_{\Delta \Omega}=6$  & sensitivity  
 \\  \hline
0.06 & 2.0 & $3.5\ 10^{-7}$ & $4.6\ 10^{-7}$ & $5.8\ 10^{-7}$ & $7.0\ 10^{-7}$ & $3.1\ 10^{-7}$ \\
  & 2.2 & $4.5\ 10^{-6}$ & $6.0\ 10^{-6}$ & $7.5\ 10^{-6}$ & $9.0\ 10^{-6}$  & $3.6\ 10^{-6}$ \\
  & 2.3 & $1.7\ 10^{-5}$ & $2.2\ 10^{-5}$ & $2.8\ 10^{-5}$  & $3.3\ 10^{-5}$  & $1.1\ 10^{-5}$ \\
  & 2.4 & $5.9\ 10^{-5}$ & $7.8\ 10^{-5}$ & $9.8\ 10^{-5}$  & $1.2\ 10^{-4}$  & $3.4\ 10^{-5}$ \\
\hline
0.38 & 2.0 & $5.4\ 10^{-8}$ & $7.3\ 10^{-8}$ & $9.1\ 10^{-8}$ & $1.1\ 10^{-7}$ & $3.1\ 10^{-7}$  \\
  & 2.2 & $7.1\ 10^{-7}$ & $9.4\ 10^{-7}$ & $1.2\ 10^{-6}$ & $1.4\ 10^{-6}$  & $3.6\ 10^{-6}$ \\
  & 2.3 & $2.6\ 10^{-6}$ & $3.6\ 10^{-6}$ & $4.4\ 10^{-6}$  & $5.2\ 10^{-6}$  & $1.1\ 10^{-5}$ \\
  & 2.4 & $9.3\ 10^{-6}$ & $1.2\ 10^{-5}$ & $1.5\ 10^{-5}$  & $1.9\ 10^{-5}$  & $3.4\ 10^{-5}$ \\
\hline 
FB & & & & & & 90\% C.L. limit \\ \hline
0.66  & 2.0 & $3.1\ 10^{-8}$ & $4.2\ 10^{-8}$ & $5.2\ 10^{-8}$  & $6.3\ 10^{-8}$  & $5.4\ 10^{-7}$ \\
\hline
\end{tabular}\par}
\caption{{\small Column 3 to 6: Normalization factors $\Phi^{D^\prime,\Gamma}_0$ for an enhanced diffuse flux. They are evaluated assuming that $n_{\Delta \Omega}=3$ to 6 events from that detected by IceCube are produced in a restricted region of circular window of 
8$^\circ$ ($\Delta\Omega=0.06$ sr) and 20$^\circ$ ($\Delta\Omega=0.38$ sr). The last row reports the prediction from a region with the extension of the Fermi bubbles (FB).
The last column shows the ANTARES sensitivities derived from the analysis on the FB regions. The 90\% C.L. result for the FB is reported in the last row.
\label{tab:diffu}}}
\end{table*}

Let us now assume that the Galactic fraction of the IceCube signal is produced in a region of angular extension $\Delta \Omega\ll 4\pi$ sr.
The signal can be observed as an enhanced diffuse flux from the corresponding sky region from a detector with angular resolution $\ll \Delta \Omega$.
This diffuse flux is characterized by a power law spectrum 
$E^{\Gamma} \Phi^{D^\prime,\Gamma} (E) = \Phi^{D^\prime,\Gamma}_0$, with the normalization factors in units of (GeV cm$^{-2}$ s$^{-1}$ sr$^{-1}$).
According to this flux, the number of IceCube HESE $n_{\Delta \Omega}$ originating within the solid angle $\Delta \Omega$ is
\begin{eqnarray}\label{eq:gad1}
n_{\Delta \Omega} &=& T \cdot \int \Phi^{D^\prime,\Gamma} (E) \cdot A_{eff}(E) \cdot dE\cdot d\Omega \nonumber \\
    &=& T {\Delta \Omega}\cdot \int \Phi^{D^\prime,\Gamma}_0  E^{-\Gamma} \cdot A_{eff}(E)\cdot dE \nonumber \\
    &=& T {\Delta \Omega} \cdot \Phi^{D^\prime,\Gamma}_0  \int E^{-\Gamma} \cdot A_{eff}(E)\cdot dE \nonumber \\
&=& T {\Delta \Omega} \cdot \Phi^{D^\prime,\Gamma}_0  \cdot {\cal D}_\Gamma \ .
\end{eqnarray}
By replacing the detector response derived in (\ref{eq:ic2}) into (\ref{eq:gad1}):
\begin{equation}\label{eq:gad2}
n_{\Delta \Omega} = {N_{IC} \cdot \Phi^{D^\prime,\Gamma}_0 \cdot {\Delta \Omega} \over  4\pi \cdot \Phi_0^{D,\Gamma} }
\end{equation}
and, correspondingly, the normalization factor for a diffuse spectrum in the region of extension ${\Delta \Omega}$ sr becomes:
\begin{equation}\label{eq:gad3}
\Phi^{D^\prime,\Gamma}_0 = \biggl({n_{\Delta \Omega} \over N_{IC}}\biggr) \cdot \biggl({4\pi \over {\Delta \Omega}}\biggr) \cdot \Phi_0^{D,\Gamma} 
\end{equation}
The symbol $\Phi^{D^\prime,\Gamma}$ is used to characterize the diffuse flux in the limited solid angle region that is enhanced with respect to the average diffuse flux $\Phi^{D,\Gamma}_0$ given in Eq. (\ref{eq:ic0}).

Table \ref{tab:diffu} (columns from 3 to 6) shows the normalization factors from Eq. (\ref{eq:gad3}), assuming $ n_{\Delta \Omega}= 3 \div 6$ HESE within two solid angle regions $\Delta \Omega=2\pi(1-\cos\theta)$ corresponding to a circular windows of  $\theta=8^\circ$ and 20$^\circ$ encompassing the Galactic Center.

\subsection{The directional cosmic neutrino flux} 

The strategy for the study of an enhanced diffuse flux must be different with respect to that for the search for point-like sources.
This latter relies mainly on the pointing accuracy of the telescope.
The background due to atmospheric neutrinos within a circular windows of $\theta\lesssim 1^\circ$ is small and this is not anymore true for larger values of $\theta$.
For the case of ANTARES \cite{anta}, with 5516 events in 1338 days, the number of atmospheric neutrinos within $\theta =8^\circ$ and $20^\circ$ corresponds to $\sim$  50 and 330 events, respectively. 

The energy spectrum from a cosmic signal (either point-like or diffuse) is expected to be harder than that of atmospheric neutrinos.
This latter, as measured by IceCube \cite{ic-atmo,ic-atmo-mu} and ANTARES \cite{anta-atmo}, depends on energy as  $\propto E^{-3.7}$. 
The signal should exceed the background above a certain energy threshold.
Thus, the discrimination between signal and background needs the use of the estimated energy of the event, similarly to the case of the search for a diffuse flux of high energy $\nu_\mu$ \cite{anta_diffu}.

An enhanced diffuse flux generated in a Galactic region observed within solid angle $\Delta \Omega$ can be seen as a \textit{directional} cosmic neutrino flux.
The \textit{directional} cosmic neutrino flux per flavor is defined as:
\begin{equation}\label{eq:gad}
\Phi^{{\Delta \omega},\Gamma} = \Phi^{D^\prime,\Gamma}\cdot \Delta \omega \quad \textrm{ units: GeV cm}^{-2} \textrm{ s}^{-1} 
\end{equation}
where $\Delta \omega$ represents the solid angle under which the experiment optimizes the search. Its minimum value, $\Delta \omega_{min}$ is related to the angular resolution of the detector. For ANTARES ($\theta^A_{res}\sim 0.4^\circ$ for the $\nu_\mu$ flavor) it is $\Delta \omega_{min}\simeq 1.5\times 10^{-4}$ sr.
The definition (\ref{eq:gad}) coincides with that of \cite{ancho,razza} when  $\Delta \omega=\Delta \Omega$.

According to (\ref{eq:gad3}) the directional cosmic neutrino flux measured under the solid angle $\Delta\omega$ overlapping the cosmic region ($\Delta \omega \subset \Delta \Omega $) is given by:
\begin{equation}\label{eq:gad3b}
\Phi^{{\Delta \omega},\Gamma} = 
\biggl({n_{\Delta \Omega} \over N_{IC}}\biggr) \cdot \biggl({4\pi \over {\Delta \Omega}}\biggr) \cdot \Phi^{D,\Gamma} \cdot \Delta \omega 
\quad 
\end{equation}
When the detector angular resolution is worse than the source size and the integral (\ref{eq:gad1}) extends over a region $\Delta \omega_{min} \supset \Delta \Omega $,
we return to the case of a point-like source, because outside 
$\Delta \Omega $ it is $\Phi^{D,\Gamma}\ll \Phi^{D^\prime,\Gamma}$. 
This is equivalent to have $\Delta \omega = \Delta \Omega$ in Eq. (\ref{eq:gad3b}), which thus reduces to (\ref{eq:gal5}). 

In the search for a directional cosmic neutrino excess using Eq. (\ref{eq:gad3b}), high-energy events must be selected to reduce the atmospheric background and a dedicated optimization of the observational solid angle $\Delta \omega$ with respect to the source extension $\Delta \Omega$ must be done.
If $\Delta \omega \ll \Delta \Omega$, the signal would be too faint. On the other hand, if $\Delta \omega \gg \Delta \Omega$ the signal would be too diluted with respect to the background.
The optimization of the size $\Delta \omega$ must be done based on the clustering dimension of the IceCube events itself or from astrophysical models.

\subsection{ANTARES sensitivities on an enhanced cosmic neutrino flux} 

ANTARES used an Artificial Neural Network to estimate the energy of the muons entering the detector for studying the two extended structures above and below the Galactic Center emitting $\gamma$-rays with a hard spectrum, the so-called Fermi bubbles (FB) \cite{anta-fb}.
Emission scenarios from both the FB and a broader halo region are considered in \cite{tay,razza} to provide a contribution to the IceCube event excess.
The angular size of the FB studied by ANTARES corresponds to $\Delta \omega =0.66$ sr.

The reported ANTARES sensitivity in terms of an enhanced diffuse flux from the FB region, assuming a $E^2 \Phi(E)$ spectrum without cutoff up to the PeV energies is $3.1\times 10^{-7}$ GeV cm$^{-2}$ s$^{-1}$ sr$^{-1}$. 
In the analysis, using 806 days livetime, 16 events were found, with an expected background of 11 events.
The derived 90\% C.L. upper limit is $E^2 \Phi^{FB}(E)=5.4\times 10^{-7}$ GeV cm$^{-2}$ s$^{-1}$ sr$^{-1}$ and it is reported in the last column and last row of Table \ref{tab:diffu}.

The ANTARES sensitivities to sky regions with different sizes overlapping the IceCube signal, and assuming different spectral indexes $\Gamma$, can be extrapolated from the sensitivity for the FB region for the $E^{-2}$ spectrum. The method described in section \ref{sez:anta} must be modified to include the following aspects:
(1) the optimization of the cut on the observed event energy assuming a $E^{-\Gamma}$ spectrum; 
(2) the optimization (size/center) of the search solid angle $\Delta \omega$. 

The sensitivity of an experiment depends on the background level. 
The optimization of the cut on the observed energy to maximize the signal-to-noise ratio for a $E^{-\Gamma}$ spectrum has a sizeable effect, because affects the number of background events (or, equivalently, the effective area for the signal). 
A first order evaluation of this effect shows that the ratios reported in the third column of Table \ref{tab:ANTA} must be multiplied by a factor $(1.2)^k$ for $\Gamma=2+0.1\times k$.
The optimization on the search solid angle do not change significantly (i.e., at a level above the $\pm 30\%$ uncertainty of the HESE normalization factor) the sensitivity for an enhanced diffuse flux. 
The background is represented by the atmospheric neutrino flux per unit of solid angle above the value of the cut on the observed energy. 
{In the case of FB analysis, it corresponds to 11 events/(0.66 sr $\cdot$ 806 days)=7.5 events/(sr $\cdot$ y). }
This ratio, neglecting the variation of the atmospheric neutrino flux on the local zenith angle and thus on the declination of the considered object, is almost independent of the size $\Delta\omega$ and of its central position.
The sensitivities obtained under the above assumptions are reported in the last column of Table \ref{tab:diffu} (with the exception of the last row). 
{A sensitivity equal to the expected flux corresponds to a number of signal events equal to the background and is indicative of the level of the rejection that ANTARES can set on the models based on the HESE sample. }

{Let us, as an example, quantitatively derive the ANTARES livetime necessary for an  observation at five standard deviations for the case of $\Gamma=2.3$ (2.4) and $n_{\Delta\Omega}=$6 (4) in a region of $\Delta \Omega=0.06$ sr. The normalization factor corresponds from Table \ref{tab:diffu} to $3.3\times 10^{-5}$ ($7.8\times 10^{-5}$) GeV cm$^{-2}$ s$^{-1}$ sr$^{-1}$.  
For a spectral index $\Gamma=2.3$ (2.4) the number of background events in the signal region (according to the value reported above for $\Gamma=2.0$ in the FB region) is $n_{bck}$=0.78 (0.93) y$^{-1}$.
The number of signal events per year, $n_{ANT}$, is given by folding Eq. \ref{eq:gad} with the ANTARES effective area. For the case of $\Gamma=2.3$ (2.4), it is $n_{ANT}$=2.2 (2.0) y$^{-1}$. With the considered background, a 5 standard deviations signal can be obtained in 4.0 (5.8) years livetime. 
Considering that ANTARES is running with its full configuration since 2008 with a duty cycle larger than 0.8, any model that assumes more than 2 HESE within $\Delta\Omega =0.06$ sr yields an observation larger than 2.0 (3.8) standard deviations for $\Gamma= 2.0$ (=2.4). This is valid if the search solid angle $\Delta \omega$ is exactly coincident with the signal region $\Delta  \Omega$.}

The above considerations contradict the conclusion reported in \cite{ancho}, where the normalization factors for a directional neutrino flux with different spectral indexes $\Gamma$ have been derived. 
They conclude that (for a $E^{-2}$ spectrum) \textit{the flux required to explain IceCube data is safely two orders of magnitude below the current ANTARES bound}.
This statement is a consequence of a wrong computation of the normalization factors reproduced in Table 3 and 4 of \cite{ancho}. 
Their normalization factors (for instance, that denoted as $\varphi_0^{E_\nu< 1 PeV}$) for a directional cosmic neutrino flux from an observation solid angle $\Delta \omega=\Delta \Omega$ are based  on the incorrect relation: 
\begin{equation}\label{eq:wrong}
\varphi_0^{E_\nu< 1 PeV} = \biggl( {n_{\Delta\Omega} \over n_{IC} }\biggr) \cdot \Phi_0^{E_\nu< 1 PeV} \cdot \Delta\Omega 
\textrm{, GeV cm}^{-2} \textrm{ s}^{-1}  \ .
\end{equation}
They used $n_{IC}=16$ and $\Phi_0^{E_\nu< 1\ PeV}$ as reported in their Table 2 (for $\Gamma=2$ it is $1.66\times 10^{-8}$ GeV cm$^{-2}$ s$^{-1}$ sr$^{-1}$) derived from \cite{ic1}. Then, $n_{\Delta\Omega} =4$ and ${\Delta\Omega} =0.06$ sr are used for Table 3 and $n_{\Delta\Omega} =6$ and ${\Delta\Omega} =0.38$ for Table 4.
By comparing Eq. (\ref{eq:wrong}) with Eq. (\ref{eq:gad3b}), it is possible to observe that the normalization factors in Table 3 and 4 of \cite{ancho} are wrong by a factor $(4\pi/\Delta\Omega)$, i.e., the first column of Table 3 must be multiplied by 209 and that of Table 4 by a factor of 33.  
When multiplied for these factors, the values reported in the two mentioned Tables are in agreement with that reported for the similar case in this paper. 

\section{Discussion and perspective for a Mediterranean detector}\label{sez:disc}

The ANTARES detector has sufficient sensitivity to test different models that explain a fraction of the IceCube HESE sample in terms of a Galactic component.

First, in \cite{pr} it was assumed that each individual IceCube event could be associated with a known source.
The correctness of such a hypothesis can probably be tested only using upgoing CC $\nu_\mu$ interactions, if the muon direction is measured with a sufficient precision to unambiguously associate the source.  
Objects located in the Southern sky are thus visible in the Northern hemisphere, where the ANTARES telescope is located.
The predicted flux for each individual source in \cite{pr} ranges between $(0.6 - 1.7)\times 10^{-8}$ GeV cm$^{-2}$ s$^{-1}$, in some case very close to the 90\% C.L. upper limits already published by ANTARES (Fig. 3 of \cite{anta}).
In particular, for three of the considered objects: PKS 2005-489 (associated with the IceCube event 12), MGRO J1908+06 (associated with event 33) and H 2356-309 (associated with event 10) the  90\% C.L. limits correspond to 1.39, 2.32 and 2.35$\times 10^{-8}$ GeV cm$^{-2}$ s$^{-1}$, respectively.
These limits hold for a $E^{-2}$ flux.
 
For a softer energy spectrum, ANTARES constrains more severely the model: in the prediction, the normalization factors would increase as in the third column of Table \ref{tab:IC} (i.e., a factor 48 for $\Gamma=2.3$), while the ANTARES upper limits increase as the third column of Table \ref{tab:ANTA} (i.e., a factor 21 for $\Gamma=2.3$).
However, a softer spectrum seems incompatible for most of the sources considered in \cite{pr} with the extrapolated SED obtained from $\gamma$-ray observations.

Second, the hypothesis that different IceCube events originate from the same object (hot spot) has been tested, with the assumption of a power law spectrum with $\Gamma$ ranging from 2.0 to 2.5.
Depending on the number $n_p$ of events in the hot spot, the normalization factors are reported in Table \ref{tab:point}.
The ANTARES 90\% C.L. upper limit obtained from studying the Galactic Center region excludes that a single point-like source produces more than 5 IceCube events, assuming a spectral index $\Gamma=2.0$. 
The limit excludes a single point-like source yielding a cluster of more than 2 events for $\Gamma=2.3$, while the presence of a cluster made of two or more events is excluded for  $\Gamma>2.3$.

The third considered hypothesis concerns the possibility that a clustering of events is produced in a limited region in (or near) the Galactic Plane. 
This yields an enhanced diffuse neutrino flux, Eq. (\ref{eq:gad3}), whose intensity depends on the region solid angle $\Delta \Omega$, and on the number of events $n_{\Delta \Omega}$ belonging to the cluster.
This enhanced diffuse neutrino flux can be observed as a directional excess, Eq. \ref{eq:gad3b}.
Predictions for the normalization factor of the enhanced diffuse flux for different values of $\Delta \Omega$ and $n_{\Delta \Omega}$ are reported in Table \ref{tab:diffu}, assuming an unbroken power law with different spectral indexes $\Gamma$.

The ANTARES collaboration has produced results from data collected from 2008 to 2011 related to a wide sky area below and above the Galactic Plane, the Fermi bubbles. 
The values of the ANTARES sensitivities around the HESE Galactic hot spot for an enhanced diffuse flux, derived from the analysis \cite{anta-fb} and for different spectral indexes $\Gamma$, are reported in the last column of Table \ref{tab:diffu}.
According to these values, a dedicated search for a directional neutrino flux around the IceCube hot spot would produce a positive result for any spectral indexes $\Gamma \ge 2.0$, if $\Delta \Omega\le 0.06$ sr (or circular window of $\theta<8^\circ$) and $n_{\Delta \Omega}>2$. 
For a signal spread out on a larger circular window, the minimum sensitivity would correspond to  a higher $n_{\Delta \Omega}$.
 


ANTARES will continue to take data at least until the end of 2015.
In parallel, it should be mentioned that Phase 1 of KM3NeT \cite{km3} plans to deploy 8 towers and about 30 strings by 2016. 
This new Northern infrastructure for neutrino detection will have an effective area for the $\nu_\mu$ flavor a factor of 3-4 times larger than that of ANTARES, with similar angular resolution for the $\nu_\mu$ direction.
Within a few years of operation, this first stage towards a cubic kilometer detector, eventually combining the results with that of  ANTARES, will have enough sensitivity to confirm or exclude sources with fluxes at a level of that reported in \cite{pr} and in the field-of-view of the Mediterranean Sea. 
The sources outside the field-of-view of ANTARES and KM3NeT can be tested using upward-going muons by IceCube itself.

\begin{acknowledgments}

\noindent I would like to thank J. Brunner, A. Kouchner and F. Vissani for useful discussions, comments, criticism and suggestions that significantly improved this work. I thanks also many members of the ANTARES and KM3NeT Collaboration for comments and in particular S. Cecchini, P. Coyle, R. Coniglione, V. Kulikovskiy and A. Margiotta. 
\end{acknowledgments}

\end{document}